\documentstyle[12pt]{article}

\def\beqa{\begin{eqnarray}}
\def\eeqa{\end{eqnarray}}
\def\beq{\begin{equation}}
\def\eeq{\end{equation}}

\def\ad{\dot{a}}

\def\gu{g^{\mu\nu}}
\def\gd{g_{\mu\nu}}

\def\unu{^{\nu}}
\def\dmu{_{\mu}}  
\def\dnu{_{\nu}}

\def\dmunu{_{\mu\nu}}

\def\ddemu{_{;\mu}}  
\def\ddenu{_{;\nu}}

\def\pa{\partial}

\def\bib#1{$^{\ref{#1}}$}

  \let\Lam=\Lambda

\let\gam=\gamma

\def\pr{{\it Phys. Rev.}\ }
\def\prl{{\it Phys. Rev. Lett.}\ }
\def\pl{{\it Phys. Lett.}\ }
\def\np{{\it Nucl. Phys.}\ }
\def\modpl{{\it Mod. Phys. Lett.}\ }
\def\ijmp{{\it Int. Journ. Mod. Phys.}\ }

\def\cqg{{\it Class. Quantum Grav.}\ }

\def\grg{{\it Gen. Relativ. Grav.}\ }

\def\apj{{\it Ap. J.}\ }
\def\aa{{\it Astron. Astrophys.}\ }

\def\rmp{{\it Rev. Mod. Phys.}\ }

\def\ie{{\it i.e. }}
\def\eg{{\it e.g. }}

\def\p{\varphi}

\def\P{\bf\Phi}
\def\l{\cal L}
\begin{document}
\def\bib#1{[{\ref{#1}}]}
\begin{titlepage}
	 \title{String Dilaton  Fluid Cosmology}

 \author{{S. Capozziello, G. Lambiase, and R. Capaldo}
\\ {\em Dipartimento di Scienze Fisiche "E.R. Caianiello",}
\\ {\em  Universit\`{a} di Salerno I-84081 Baronissi (SA) Italy,}
\\{\em Istituto Nazionale di Fisica Nucleare, Sez. di Napoli }
\\{\em Mostra d'Oltremare Pad. 20, I-80125 Napoli, Italy.}}
	      \date{}
	      \maketitle

	      \begin{abstract}
\noindent We investigate $(n+1)$-dimensional string-dilaton 
cosmology  with effective dilaton potential in presence 
of perfect-fluid matter.
We get exact solutions parametrized by the constant $\gam$
of the state equation 
$p=(\gam-1)\rho$, the spatial dimension
number $n$, the bulk of matter, and the spatial curvature constant 
$k$. 
Several interesting cosmological behaviours 
are selected. Finally we  discuss the recovering of ordinary Einstein gravity
starting from string dominated regime and a sort of asymptotic freedom
due to string effective coupling.
	      \end{abstract}

\vspace{20. mm}

\vspace{20. mm}
e--mail addresses:\\
capozziello@vaxsa.csied.unisa.it\\ 
lambiase@vaxsa.csied.unisa.it
\vspace{10. mm}

	      \vfill
	      \end{titlepage}

\section{\normalsize \bf Introduction}

The goal to connect the large scale structure of the universe 
with fundamental particle physics is one of the 
major issues of modern cosmology. One should  be able to relate
the great amount of recent observations (COBE, HST, surveys of galaxies, etc.
see for example \bib{cobe}--\bib{bartelmann}) 
with some fundamental theory following the scheme
firstly drawn by Sakharov \bib{sakharov}:
the primordial quantum fluctuations should have been enlarged 
to astronomical
sizes by some expansion mechanism (\eg inflation) and then give rise
to galaxies, clusters and superclusters of galaxies, when the perturbations
grow away from the cosmological background \bib{kolb}. However, there
are some shortcomings in this approach since
a very fundamental theory of gravity, connected to the 
other interactions, does not 
exist yet (\ie theories exist but none of them has been, till now,
experimentally tested).

All the unification schemes are defective for several reasons when
they have to take into account the particle families actually discovered.
The standard model of particles (actually detectable) is in agreement
with  cosmological observations 
below energies of 1 TeV, while the unification
theories require  energies from $10^{14\div 16}$GeV of grand unified theories
(GUT) to
$10^{19}$GeV of full quantum gravity. Such huge energies are 
not detectable by the today facilities.

Furthermore, all the present inflationary models have to be fine
tuned in order to get perturbation spectra able to reproduce the
observed microwave background isotropy and to explain the large scale
structures \bib{guth}. 

This situation is unsatisfactory, but some progress have been done in recent
years and, very likely, the only available way to understand the large
scale structures in the universe and to attempt a superunification with
observable consequences is to investigate theories which, at the same
time, explain the main riddles of astrophysics (large--scale structures
and dark matter) and of high energy physics (hierarchy problem, parity
violation, number of families etc.) under the same {\it astroparticle}
standard.

For example, let us consider the inflationary paradigm.
We can change our point of view and taking into consideration
the geometric side (without considering matter scalar fields)
\bib{caianiello}. In such a way, we can construct classes 
of cosmological models
which avoid the initial singularity, 
evolve towards the decelerating state of standard cosmology, and
can be connected with the perturbation spectra without fine tuning.
Alternatively, 
the so called {\it extended} and {\it hyperextended} inflations
\bib{la} solve the shortcomings of $old$ and $chaotic$ inflations
by changing the gravitational sector through a nonminimal coupling. 
These models allow to complete
the {\it graceful exit} from the {\it false} toward the {\it true} 
vacuum,  avoid 
the overwhelming presence of topological
defects after the phase transition, yield  realistic power spectra
for cosmological perturbations. Besides, all unification theories
forecast, in the low energy limit, nonminimal couplings between 
matter fields and geometry (in particular when a quantum field theory is
built on  curved spacetimes \bib{birrell}); then it seems reasonable
to invoke a variation of Newton {\it constant} $G_{N}$ toward  the 
early epochs from both astrophysical and particle sides.
Furthermore, the {\it cosmological constant} $\Lambda$
has to vary from the today observed upper limit 
$(\Lambda \simeq 10^{-56}cm^{-2})$ \bib{kolb},
 in order to allow
one or more than one phase transitions. In general, a lot of 
theories follow this paradigm: the universe has
a fundamental quantum--mechanical nature in which gravity strictly depends
on some scalar fields which have to lead the system toward the observed
situation $(G_{eff}\rightarrow G_{N},\, \Lambda_{eff}\rightarrow\Lambda)$
at present epoch \bib{adler}.

String theory can be included in this set of theories. 
It is one of the most serious attempts, in the last thirty
years, to get the great unification, since it avoids the shortcomings of
quantum field theories essentially due to the pointlike nature of particles
(renormalization) and includes gravity in the same conceptual scheme of other
fundamental interactions (the graviton is just a string $mode$ as the other
gauge bosons).
 In general, all the elementary particles can be interpreted as different
modes of a single string whose oscillation frequencies determine
the particle masses.

In the low
energy limit (that is  far from the Planck energy scale), a string
dominated universe is described by the {\it tree--level effective action}
\bib{scherk},\bib{veneziano},\bib{dealwis},\bib{green} in which only massless
modes appear {\it (zero modes)}. 
From it, the lowest order string $\beta$-function 
equations can be derived. These equations, for the closed string, describe 
the dynamics of
free long range fields which are: the scalar mode (the dilaton $\p$), 
the Kalb--Ramond field strength
$H_{\mu\nu\lambda}$ (usually called the "torsion field"), and the graviton. 
The last two ones are tensor modes.
These three fields are nonminimally coupled in the 
string effective Lagrangian. 
In addition, there is a constant related to
the central charge of the string theory which vanishes for the critical
number of dimensions 10 or 26.
In general, one can consider strings propagating in a universe of spatial
dimension $n$ \bib{veneziano},\bib{green}. If $n$ is not the  
critical dimension (that is
$n< 26$ for bosonic strings or $n< 10$ for superstrings), we could have one of
the two possibilities: either we have some $n_{c}$ of compact internal 
dimensions 
which make up the rest of the spatial dimensions which we take to be static
and to correspond to conformal theories on the world--sheet
\bib{green},
or we are dealing with a noncritical theory. The difference
between the two approaches results in the low energy limit action where
we have to consider the constant
\beq
\label{charge}
\Lambda=\frac{2}{3}(d_{c}-n-n_{c})\,,
\eeq
where $d_{c}$ is the critical spatial dimension. If the theory is critical,
we have $\Lambda=0$, if it is noncritical $\Lambda\neq 0$.
Usually, the $n$--dimensional space is taken to be a box of lengths 
$a_{i}$. In cosmology, that  is in the low
energy limit, such lengths are the scale factor(s) of an (an)isotropic
universe \bib{tseytlin}. 

In any case, string theory is one of the unification scheme able to describe
all the fundamental interactions under the same standard \bib{green},
so the string--dilaton cosmology could be the way to solve the above
astrophysical riddles.  
However, the fact that only the massless excited
states are used suggests that the effective action is not a valid description
for probing the highest energies associated with string theory but
we may hope that, through the $\beta$-function equations, we are investigating
physics associated with events from  the string scale down to the
GUT scales and below.

On the other hand, cosmology is the only 
available way to test the consequences of the theory due
to the high energies requested by every quantitative prediction for
fundamental strings.
The today detectable {\it remnants} of primordial processes
could be a test to which  to compare the theory
and a lot of open questions of astrophysics, like dark matter
and large scale structure, could be solved by strings and their dynamics
\bib{tseytlin}.
 
In a series of papers, it has been shown that it is possible to obtain
exact solutions for string--dilaton cosmology in  homogeneous 
 spacetimes 
(see for example \bib{tseytlin}--\bib{copeland}).
Most of them possess the string major feature, the {\it duality}, so that
when $a(t)$, the scale factor of the universe, is a solution of the 
equations of motion, $1/a(t)$ is a solution too
(in Planck's units). In other words, the whole
spacetime behaves like a string\footnote{The string duality property can
be stated as follows: a (closed) string moving on a circle of radius $r$
is equivalent to one moving on a circle of radius $\lambda^{2}_{s}/r$
where $\lambda_{s}^{2}=\alpha'\hbar$ is connected to the fundamental
length of the string and $\alpha'$ is a constant called 
the {\it Regge slope}. 
From this fact,  strings {\it do not see} scales
smaller than their proper natural scale.}
This fact implies that the universe could have a minimal size comparable
with the string scale and a maximal one comparable with
the inverse of such a scale.
In other words,  physical size  and 
 initial singularity  of the universe could be connected to the 
features of string theory
and if the dynamics of early universe is governed by a string gas, 
 it does not begin from a null size but from a finite size
that is the Planck scale.
In order to determine such a dynamics, string cosmology needs
 the existence of two kinds of fields, which are $dual$ between them.
At small scales, one of them is nonmassive and dominates the dynamics;
at large scales, it is the other one which becomes dominant and nonmassive. 
The first contribution comes from the  dilaton 
which leads
 the expansion of observed spatial dimensions; the other
one is due to the {\it graviton}, that is to the  geometry. 

A fundamental consequence of such a situation is that the gravitational
interaction results nonminimally coupled with dilaton; then the effective
string theory is a sort of Jordan--Brans--Dicke theory, \ie an induced gravity
theory \bib{tseytlin},\bib{pls},\bib{dual}. 
We have to note that general relativity is not invariant under such
a duality transformation.

Furthermore, in string--dilaton cosmology, there exist the
possibility that inflation can occur without relying on a potential energy 
density. However, the dilaton potential has to be suitable for gaugino
condensation. In this case, we need a mechanism (not yet found) able
to explain the vanishing of the cosmological constant \bib{weinberg}.

From this point of view,  the presence of 
a potential for the dilaton and the fact that the string effective 
action has to
contain matter sources are crucial ingredients. The sources may be represented
in a perfect fluid form, but always with an equation of state which
consistently follows from the solution of the string equations of motion in 
the given cosmological background. This description, probably, does not
work in  the very high curvature regime but it is necessary in order
to recover (even through inflation) 
the standard cosmological picture which we today observe.

In this paper, we want to address this problem: we deal with a 
$(n+1)$--cosmological string effective action in which 
perfect fluid matter is present
and we want to investigate how it affects string background dynamics.
The method is the same which is used in \bib{qua} and
in \bib{bra}.
The dilaton potential is also considered and it is possible to show that
in a $(n+1)$ Friedman--Robertson--Walker (FRW) model, the
number of spatial dimension $n$, the amount of ordinary matter,
the  sound speed in the state equation 
(given by a constant $\gam$), and such a potential are strictly related.
The results are easily extensible to any Bianchi model.

In Sec.2, we write down and discuss the 
string--dilaton--fluid matter effective action showing that it can
be derived from the general action of a nonminimally coupled theory of 
gravity. 
Sec.3 is devoted to
the derivation of equations of motion. In Sec.4, we discuss some realistic
forms of dilaton potential.
Sec.5 is devoted to the exact solutions and to
their cosmological classification. In Sec.6, we discuss the results and
draw conclusions.

\section{\normalsize \bf The string--dilaton--fluid matter effective action}

The most general  action of a scalar field $\psi$ nonminimally coupled with 
the Ricci scalar $R$ in $(n+1)$ dimensions is 
\beq
\label{1}
{\cal A}=\int 
d^{n+1}x\sqrt{-g}\left[F(\psi)R+\frac{1}{2}\gu\psi\ddemu\psi\ddenu
-W(\psi)+\tilde{\l}_{m}\right]\,,
\eeq
where $F(\psi)$ is the coupling, $W(\psi)$ is the potential for the field
$\psi$, $\tilde{\l}_{m}$ is the ordinary matter contribution,
and $\mu,\nu= 0,...,n$. The string--dilaton effective action 
(\ie the string action in the low energy limit) containing the contributions
due to the dilaton $\p$ and to the graviton $\gd$ (disregarding 
the antisymmetric torsion field\footnote{Actually, as shown
in \bib{copeland}, such an antisimmetric field can place a lowerbound on the 
allowed values of the dilaton and can give rise to a radiation dominated
FRW universe. However, the main role in  dynamics is played by $\p$,
 $\gd$ and  the string--loop interactions \bib{damour}.}) 
is easy recovered by the 
transformations
\beq
\label{2}
\psi=\exp[-\p]\,,\;\;\;F(\psi)=\frac{1}{8}\exp[-2\p]\,,\;\;\;
W(\psi)=U(\p)\exp[-2\p]\,,
\eeq
which specify the coupling. The action (\ref{1}) becomes
\beq
\label{3}
{\cal A}=\int 
d^{n+1}x\sqrt{-g}\left\{\exp[-2\p]\left[R+4\gu\p\ddemu\p\ddenu
 - U(\p)\right]+\tilde{\l}_{m}\right\}\,.
\eeq
We  have supposed the  fluid--matter 
contribution  $\tilde{\l}_{m}$ minimally
coupled with string--dilaton degrees of freedom, and, in principle,
we have to specify it with respect to the other fields. 
Then we take into consideration a $(n+1)$ dimensional 
FRW metric since we want to deal with the cosmological
problem. In this situation, the Lagrangian in (\ref{1}) 
becomes pointlike and we get  
\beq
\label{4}
{\l}=n(n-1)a^{n-2}{\ad}^{2}F(\psi)+2n\ad a^{n-1}\dot{\psi}F'(\psi)-
n(n-1)ka^{n-2}F(\psi)+\frac{1}{2}a^{n}\dot{\psi}^{2}-a^{n}W(\psi)+
a^{n}\tilde{\l}_{m}\,,
\eeq
while the Lagrangian in (\ref{3}) becomes
\beq
\label{5}
{\l}=\frac{1}{8}a^{n}e^{-2\p}\left[n(n-1)\left(\frac{\ad}{a}\right)^{2}-
4n\dot{\p}\left(\frac{\ad}{a}\right)-n(n-1)\frac{k}{a^{2}}+4\dot{\p}^{2}
-8U(\p)\right]+a^{n}\tilde{\l}_{m}\,.
\eeq
The constant $k$ is the spatial curvature constant which can be
$k=\pm 1,0$ for spatially closed, open and flat cosmological models,
respectively.

The pointlike Lagrangians (\ref{4}) and (\ref{5}) can be 
easily extended to any anysotropic and homogeneous 
spacetime \bib{any},\bib{lambdat}. Here, for the sake of simplicity,
we shall treat only FRW models. 

Now we have to determine the fluid matter
contribution. In this case, the stress--energy tensor
has the form
\beq
\label{6}
T\dmunu=(p+\rho)u\dmu u\dnu-\gd p\,,
\eeq
where $p$ and $\rho$ are, respectively, the pressure and energy--matter
density.
It has to satisfy the contracted Bianchi identities
\beq
\label{7}
T\dmu{\ddenu}\unu=0\,.
\eeq
In  FRW $(n+1)$--dimensional spacetimes, (\ref{7}) becomes
\beq
\label{8}
\dot{\rho}+nH(p+\rho)=0\,,
\eeq
where ${\displaystyle H=\frac{\ad}{a}}$ is the Hubble parameter.
Imposing the usual equation of state
\beq
\label{09}
p=(\gam-1)\rho\,,
\eeq
where $\gam$ is a constant related with the sound speed, we get 
\beq
\label{9}
\rho=Da^{-n\gam}\,.
\eeq  
$D$ is a positive integration constant related to the matter density at
$t=t_{0}$ \bib{qua},\bib{bra}.  

As we said, $\gam$ determines the thermodynamical state of matter
fluid. In standard Friedman--Einstein cosmology, we have, for example,
$\gam=1$ for  dust, $\gam=4/3$ for  radiation, $\gam=2$ for stiff matter
and $\gam=0$ for scalar field matter. The last case is particular since
yields a repulsive gravity. It is found in inflationary epoch.
This scheme takes into consideration only stationary states  but it
does not consider
the phase transitions between one regime to another. Actually, we
should take into consideration a sort of step--function which,
at equilibrium, assumes certain values.
In $(n+1)$--dimensional
string--dilaton cosmology, as we shall see below, $\gam$ can depend
on $n$ and, for any dimension, it can assume different values which
means a different thermodynamical interpretation.

\vspace{2. mm}  

The scale--factor duality symmetry is present in the lowest order 
string effective action and, as we said, it
means that the transformation of  scale factor of a homogeneous and
isotropic target space metric, $a(t)\rightarrow a^{-1}(t)$, leaves
the model invariant, provided that the dilaton field is transformed
as
\beq
\label{10}
{\P}=\p-\frac{n}{2}\ln a\,.
\eeq
Finally, taking into account the matter contribution (\ref{9}) and
the duality transformation (\ref{10}), Lagrangian  (\ref{5}) becomes
\beq
\label{11}
{\l}=\frac{1}{2}e^{-2\P}\left[4\dot{\P}^2-n\left(\frac{\ad}{a}\right)^{2}
-n(n-1)\frac{k}{a^{2}}-8V(\P(\p,a))\right]+Da^{n(1-\gam)}\,.
\eeq
where the potential $U(\p)\rightarrow V({\P(\p,a)})$.
However, the duality invariance strictly depends on the form of the potential
$V$ and on $k$.
 Lagrangian (\ref{11}) can be furtherly simplified introducing the 
variable 
\beq
\label{12}
Z=\ln a\,,
\eeq
so that
\beq
\label{13}
{\l}=\frac{1}{2}e^{-2\P}\left[4\dot{\P}^2-n\dot{Z}^{2}
-n(n-1)ke^{-2Z}-8V(\P,Z))\right]+De^{n(1-\gam)Z}\,.
\eeq 
We have to note that in such a new variables the duality invariance
becomes a sort of parity invariance since $Z$ and $-Z$ are solutions
\bib{cimento},\bib{dual}.

\section{\normalsize \bf The equations of motion}

Fom (\ref{13}), the Euler--Lagrange equations are
\beq
\label{14}
n\ddot{Z}-2n\dot{\P}\dot{Z}+n(n-1)ke^{-2Z}-4\frac{\pa V}{\pa Z}+
n(1-\gam)De^{n(1-\gam)Z}e^{2\P}=0\,,
\eeq
and
\beq
\label{15}
\ddot{\P}-\dot{\P}^{2}-\frac{n}{4}\dot{Z}^{2}-\frac{n(n-1)}{4}ke^{-2Z}-2V+
\frac{\pa V}{\pa \P}=0\,.
\eeq
The first one is the Friedman--Einstein equation, the second one is the
Klein--Gordon equation.
The energy function connected with (\ref{13}) is
\beq
\label{16}
E_{\l}=\frac{\pa {\l}}{\pa \dot{Z}}\dot{Z}
+\frac{\pa {\l}}{\pa \dot{\P}}\dot{\P}-{\l}\,,
\eeq
and then
\beq
\label{17}
4\dot{\P}^{2}-n\dot{Z}^{2}+n(n-1)ke^{-2Z}+8V-2De^{n(1-\gam)Z}e^{2\P}=0\,,
\eeq
which corresponds to the $(0,0)$--Einstein equation.
The dynamical system (\ref{14}), (\ref{15}), and (\ref{17}) is that we are
going to study. Actually, such a system is a parametric system depending
on the form of the potential $V$, the bulk of fluid matter $D$, 
the thermodynamical state of fluid $\gam$, the spatial
curvature $k$ and the spatial dimension $n$. We shall show that all
these parameters are connected giving specific cosmological features.

\section{\normalsize \bf The  effective potential}
As we said, the presence of the dilaton potential is a crucial ingredient
in string affective action for several reasons. From a cosmological point
of view, it is needed in order to achieve some kind of inflation 
(\eg chaotic inflation) and to solve the cosmological constant problem.
From a fundamental physics point of view, it is needed in order to
break supersymmetry taking into account the  gaugino 
condensation,
 to compensate the contributions due to bosons and fermions
in the vacuum energy, to lead the dynamics of particles like the axions
\bib{kolb}.  In general, it appears as a nontrivial combination
of exponentials. This feature expresses the fact that string--loop interactions
have an expansion in term of their coupling constants \bib{damour}.
It is related also to the string central charge as we mentioned above.
By taking into account the transformations (\ref{10}) and (\ref{12}),
the dilaton potential becomes a combination of geometrical and matter
degrees of freedom, that is of $Z$ and ${\P}$. 

Here we shall show that its dynamics is strongly connected with that of
standard fluid matter. We shall consider three representative
cases:
\beq
\label{case1}
V(Z,{\P})=\frac{D}{4}e^{n(1-\gam)Z}e^{2{\P}}\,,
\eeq
\ie the potential compensates the fluid matter contribution;
\beq
\label{case2}
V(Z,{\P})=0\,,
\eeq
\ie when the dilaton potential is absent;
\beq
\label{case3}
V(Z,{\P})=\Lambda \,,
\eeq
the constant case, which is the properly called dilaton potential.
Actually, the second case is nothing else but a subcase of the third one
when the constant $\Lambda$ is chosen to be zero or it is extremely small.

\section{\normalsize \bf The solutions and their cosmological interpretation}

Let us now discuss the various situations that can be parametrized
by $n$, $D$, $k$, $\gam$ and the form of the potential.

\subsection{\normalsize \bf The compensating potential}

In the first of above cases, we are assuming that the amount of potential 
energy
due to the dilaton  is exactly comparable with the amount of fluid matter
so that the two contributions annihilate each other. It could be interpreted
as a sort of energy conversion. The dynamical system (\ref{14}), 
(\ref{15}), and (\ref{17}) becomes
\beq
\label{14'}
n\ddot{Z}-2n\dot{\P}\dot{Z}+n(n-1)ke^{-2Z}=0\,,
\eeq
\beq
\label{15'}
\ddot{\P}-\dot{\P}^{2}-\frac{n}{4}\dot{Z}^{2}-\frac{n(n-1)}{4}ke^{-2Z}=0\,,
\eeq
\beq
\label{17'}
4\dot{\P}^{2}-n\dot{Z}^{2}+n(n-1)ke^{-2Z}=0\,.
\eeq
We have to distinguish the cases with $k=0$ and $k\neq 0$. If $n=1$,
the two cases coincide since a bidimensional theory (\ie a spatial dimension
plus time) is not affected by spatial curvature. This is true also for the 
considerations below. 

In the flat case, the general solution is
\beq
\label{sol1}
{\P}(t)=-\frac{1}{2}\ln\left|t-t_{0}\right|+{\P}_{0}\,,
\eeq
\beq
\label{sol2}
Z(t)=\pm\frac{1}{\sqrt{n}}\ln\left|t-t_{0}\right|+Z_{0}\,,
\eeq
and by using the inverse transformations of (\ref{10}) and (\ref{12}), we get
the cosmological behaviours
\beq
\label{sol3}
a(t)=a_{0}(t-t_{0})^{\pm 1/\sqrt{n}}\,,
\eeq
\beq
\label{sol4}
\p(t)=\p_{0}-\left(\frac{1\pm\sqrt{n}}{2}\right)\ln\left|t-t_{0}\right|\,.
\eeq
Clearly they are invariant for duality and the ordinary fluid matter
(\ie $D$ and $\gam$) is not present in the evolution. They are the same
as in a theory without potential (a critical or a free theory) and
without ordinary matter \bib{cimento}. 
Being ${\displaystyle \frac{1}{\sqrt{n}}}\leq 1$ in any case, the evolution
is Friedmanian for any spatial dimension.
In the case of $n=1$, the scale factor
becomes linear in $t$ while its dual solution pole--like.

If the model is not spatially flat (\ie $k\neq 0$), a 
solution is
\beq
\label{sol5}
{\P}(t)={\P}_{0}-\frac{n}{2}\ln |t-t_{0}|\,,
\eeq
\beq
\label{sol6}
Z(t)=\frac{1}{2}\ln|k|+\ln |t-t_{0}|\,,
\eeq
which means
\beq
\label{sol7}
a(t)=a_{0}\sqrt{|k|}(t-t_{0})\,,\;\;\;\;\;
\p(t)=\tilde{\p}_{0}=\p_{0}+\frac{n}{4}\ln|k|\,.
\eeq
The universe evolves linearly and the gravitational coupling is 
a constant. The curvature constant must be $k<0$, that is $k=-1$.
Considering the results we have got, the potential (\ref{case1}) can
be regarded as a sort effective cosmological constant which is,
at any time, comparable with the bulk of ordinary fluid matter
and annihilate each other.

\subsection{\normalsize \bf The constant potential}

This situation shows a lot of interesting subcases in which the amount
$D$ and the thermodynamical state $\gam$ of  ordinary matter 
strongly influence  dynamics.
The equations of motion become
\beq
\label{14''}
n\ddot{Z}-2n\dot{\P}\dot{Z}+n(n-1)ke^{-2Z}-
n(1-\gam)De^{n(1-\gam)Z}e^{2\P}=0\,,
\eeq
\beq
\label{15''}
\ddot{\P}-\dot{\P}^{2}-\frac{n}{4}\dot{Z}^{2}-\frac{n(n-1)}{4}ke^{-2Z}-
2\Lambda=0\,,
\eeq
\beq
\label{17''}
4\dot{\P}^{2}-n\dot{Z}^{2}+n(n-1)ke^{-2Z}+8\Lambda-2De^{n(1-\gam)Z}e^{2\P}=0\,.
\eeq
The critical case (\ref{case2}) is just a subcase. The summary of all
the situations is given in Tab.1.

Due to the structure of the equations of motion, the case with
$\gam=1$ deserves more attention. For $n=3$, it corresponds to a dust
dominated universe. In Tab.2, all combinations of the 
parameters in this particular case are summarized.

Before discussing the various situations, we have to stress again that 
the cases $\Lambda >0$, which is a positive defined potential,
 $\Lambda <0$, which is a negative defined potential, and $\Lambda=0$
 have all physical interest as pointed out, for example, in \bib{easther}.
As mentioned above,  
$\Lambda$ is the string central charge depending on the number of spatial
dimensions (see Eq.(\ref{charge})) and on the string--loop interactions
\bib{damour}.  At a very fundamental level, the combinations of these two 
facts can determine the sign of the effective dilaton potential changing,
from a cosmological point of view, the overall evolution of the universe.

Let us now show the various situations, that, as in Tab.1, can be
classified by $\Lambda$, $k$, and $D$. As further parameters, we use
$n$ and $\gam$ which are always related. It is interesting to compare
the cases with $D=0$ and $D\neq 0$ in order to see how the bulk of ordinary
matter influences the evolutions.

\subsubsection{\normalsize \it The cases with $D=0$}

If $k=D=0$, by solving the system (\ref{14''})--(\ref{17''}) and inverting
Eqs.(\ref{10}), (\ref{12}) we get the  solutions 
\bib{veneziano},\bib{tseytlin},\bib{cimento}
\beq
\label{solu1}
a(t)=a_{0}\left[\tan\sqrt{2\Lambda}(t-t_{0})\right]^{\pm 1/\sqrt{n}}\,,
\eeq
\beq
\label{solu2}
\p(t)=\pm\frac{\sqrt{n}}{2}\ln|\tan\sqrt{2\Lambda}(t-t_{0})|-
\frac{1}{2}\ln|\sin\sqrt{2\Lambda}(t-t_{0})|+\p_{0}\,,
\eeq
for $\Lambda>0$;
\beq
\label{solu3}
a(t)=a_{0}\left[\tanh \sqrt{2|\Lambda|}(t-t_{0})\right]^{\pm 1/\sqrt{n}}\,,
\eeq
\beq
\label{solu4}
\p(t)=\pm\frac{\sqrt{n}}{2}\ln|\tanh\sqrt{2|\Lambda|}(t-t_{0})|-
\frac{1}{2}\ln|\sinh\sqrt{2|\Lambda|}(t-t_{0})|+\p_{0}\,,
\eeq
for $\Lambda<0$ and
\beq
\label{solu5}
a(t)=a_{0}(t-t_{0})^{\pm 1/\sqrt{n}}\,,
\eeq
\beq
\label{solu6}
\p(t)=\p_{0}-\left(\frac{1\pm\sqrt{n}}{2}\right)\ln\left|t-t_{0}\right|\,.
\eeq
for $\Lambda=0$. Duality is evident.
Eq.(\ref{solu1}) shows an inflationary behaviour while (\ref{solu3})
goes towards a stationary universe for $t\rightarrow +\infty$. 
For $\Lambda=0$, the situation is completely equivalent to that above.

If $\Lam = D=0$ and $k\neq 0$, we have, also here as above,
\beq
\label{solu7}
a(t)=a_{0}\sqrt{|k|}(t-t_{0})\,,\;\;\;\;\;
\p(t)=\tilde{\p}_{0}=\p_{0}+\frac{n}{4}\ln|k|\,.
\eeq

\subsubsection{\normalsize \it The cases with $D\neq 0$}

The presence of fluid matter can strongly influence 
the cosmological evolution.
If $\Lam\neq 0$, $k\neq 0$, and $D\neq 0$, an exact solution is
\beq
\label{solu8}
a(t)=a_{0}\cos\sqrt{|k|}(t-t_{0})\,,\;\;\;\; 
\p(t)=\tilde{\p}_{0}=\p_{0}+\frac{1}{2}\ln\left[\frac{n|k|}{D}\right]\,.
\eeq
The constant $\Lam$ has to be
\beq
\label{solu9}
\Lam=\frac{1}{8}n(n+1)|k|>0\,,
\eeq
with $\gam=0$ and $k=-1$. This is an oscillating universe where the
gravitational coupling is constant and depend on the bulk of fluid matter.

If $\Lam =0$, $k\neq 0$, and $D\neq 0$, we have
\beq
\label{solu10}
a(t)=a_{0}(t-t_{0})\,,\;\;\;\; 
\p(t)=\p_{0}+\left(\frac{n\gam-2}{2}\right)\ln|t-t_{0}|\,,
\eeq
with $\gam< 2/n$, $k=-1$, and $n\neq 1$. The evolution is linear
and the number of spatial dimensions is connected with the sound speed of 
fluid matter, that is the thermodynamical state is determined by the
dimensionality.

When $\Lam=k=0$ and $D\neq 0$, we have
\beq
\label{solu11}
a(t)=a_{0}\left[\frac{D(t-t_{0})-\sqrt{n}}{D(t-t_{0})+\sqrt{n}}
\right]^{\pm 1/\sqrt{n}}\,,
\eeq
\beq
\label{solu12}
\p(t)=-\left(\frac{\pm\sqrt{n}-1}{2}\right)\ln |D(t-t_{0})-\sqrt{n}|-
\left(\frac{\pm\sqrt{n}+1}{2}\right)\ln |D(t-t_{0})+\sqrt{n}|+\p_{0}\,,
\eeq
but it has to be $\gam=1$.

For $\Lam\neq 0$, $k=0$, and $D\neq 0$, we have
\beq
\label{solu13}
a(t)=a_{0}\left[\cos 2\sqrt{\Lambda}(t-t_{0})\right]^{\pm 1/\sqrt{n}}\,,
\eeq
\beq
\label{solu14}
\p(t)=-\left(\frac{1\pm\sqrt{n}}{2}\right)
\ln\left|\cos 2\sqrt{\Lambda}(t-t_{0})\right|+\p_{0}\,,
\eeq
when $\Lam>0$ and
\beq
\label{solu15}
a(t)=a_{0}\left[\cosh 2\sqrt{|\Lambda|}(t-t_{0})\right]^{\pm 1/\sqrt{n}}\,,
\eeq
\beq
\label{solu16}
\p(t)=-\left(\frac{1\pm\sqrt{n}}{2}\right)
\ln\left[\cosh 2\sqrt{|\Lambda|}(t-t_{0})\right]+\p_{0}\,,
\eeq
when $\Lam<0$. The bulk of fluid matter can be any but
\beq
\gam=\frac{\sqrt{n}\pm 1}{\sqrt{n}}\,.
\eeq
A particular treatment deserves this case if $\gam=1$.
In fact we have
\beq
\label{solu17}
a(t)=a_{0}\left[\frac{D\tan[\sqrt{2\Lam}(t-t_{0})]-c_{1}}{
D\tan[\sqrt{2\Lam}(t-t_{0})]-c_{2}}\right]^{\pm 1/\sqrt{n}}\,,
\eeq
\beq
\label{solu18}
\p(t)=-\frac{1}{2}\ln\left|c_{3}-c_{4}\sin[2\sqrt{2\Lambda}(t-t_{0})]\right|
\pm\frac{\sqrt{n}}{2}
\ln\left|\frac{D\tan[\sqrt{2\Lam}(t-t_{0})]-c_{1}}{
D\tan[\sqrt{2\Lam}(t-t_{0})]-c_{2}}\right|^{\pm 1/\sqrt{n}}\,,
\eeq
where
\beq
c_{1,2}=\sqrt{D^{2}+8\Lam n\dot{Z}^{2}_{0}}\pm 
\dot{Z}_{0}\sqrt{8\Lam n}\,,\;\;\;
c_{3}=\frac{D}{8\Lambda}\,,\;\;\;
c_{4}=\frac{\sqrt{D^{2}+8\Lambda n \dot{Z}^{2}_{0}}}{8\Lambda}\,,
\eeq
if $\Lam>0$, and
\beq
\label{solu19}
a(t)=a_{0}\left[\frac{d_1 e^{2\sqrt{2\Lam}(t-t_{0})}-d_{2}}{
d_1 e^{2\sqrt{2\Lam}(t-t_{0})}-d_{3}}\right]^{\pm 1/\sqrt{n}}\,,
\eeq
\beq
\label{solu20}
\p(t)=-\frac{1}{2}\ln\left|d_{1}\sinh[2\sqrt{2\Lambda}(t-t_{0})]-D\right|
\pm\frac{\sqrt{n}}{2}
\ln\left|\frac{d_1 e^{2\sqrt{2\Lam}(t-t_{0})}-d_{2}}{
d_1 e^{2\sqrt{2\Lam}(t-t_{0})}-d_{3}}\right|^{\pm 1/\sqrt{n}}\,,
\eeq
where
\beq
d_{1}=\sqrt{8|\Lam |n\dot{Z}^{2}_{0}-D^{2}}\,,\;\;\;
d_{2,3}=D\pm\dot{Z}_{0}\sqrt{8|\Lam|n}\,.
\eeq
if $\Lam<0$. $\dot{Z}_{0}$ is an integration constant.

\section{\normalsize \bf Discussion and conclusions}
In this paper, we have discussed the string--dilaton cosmology
for a FRW spacetime in presence of minimally coupled perfect fluids.
Such  fluids are needed if we want to
connect string cosmology with standard Einstein cosmology.
As shown above the presence of fluids significantly modifies the evolution of
the scale factor and the dilaton and, in some cases, it is connected with the 
dimensionality of the string background.
However, the main role in the evolution is played by $\Lambda$ which 
determines the nature of solutions. We get power law behaviours for
$\Lambda=0$, oscillating behaviours for $\Lambda>0$, and hyperbolic ones for
$\Lambda<0$. Since $\Lambda$ is connected to the dimensionality
(see Eq.(\ref{charge})), dynamics of compactification should leave a specific 
imprint in the today observed evolution. The fluid matter plays a role 
in the rapidity by which the cosmological system evolves since it 
modulates the slope of $a(t)$.

Another important point has to be stressed.
The effective gravitational constant, for a nonminimally coupled theory as
that in (\ref{1}), can be defined, in Planck units, as
\beq
G_{eff}=-\frac{1}{2F(\psi)}\,.
\eeq
To recover the  Newton constant, it has to be $F(\psi)=-1/2$.

By using the choices (\ref{2}) and specifying the action as in (\ref{3}),
we have that 
\beq
G_{eff}=-4\exp[2\p]\,,
\eeq
and
\beq
\frac{\dot{G}_{eff}}{G_{eff}}=2\dot{\p}\,.
\eeq
This fact means that in various cases (see for example the last three ones)
it is the bulk of fluid matter (\ie $D$) which modulates the variation of
the gravitational coupling and then the strength of gravity. 
Furthermore, for several solutions, when
$t\rightarrow \pm\infty$, the gravitational coupling disappears (see \eg
all the solutions in which $\p(t)$ diverges toward $\pm\infty$).
Then, by string cosmology, we can recover, in some cases, the standard 
gravity toward
present epoch $(G_{eff}\rightarrow G_{N}$) and an asymptotically free
theory toward the past or toward the future (see for example
Eq.(\ref{solu12})) \bib{asy}. The fact that $G_{N}$ can be
essentially modulated by $D$, $\Lambda$, and $\gam$ means that the today 
observed gravitational interaction depends upon very fundamental
primordial processes which are connected with the dimensionality,
the string--loop interactions ($\Lambda$),  the thermodynamical state of 
matter ($\gam$), and the effective content of fluids in the universe 
($D$)\bib{matter}. 
All these parameters can be related to the cosmological observables in
order to reproduce the experimental limits on $G_{N}$ variation \bib{dickey}.
In a forthcoming paper, we will face this task.

\vspace{3. mm}

\noindent{\bf Acknowledgments}\\
The authors are grateful to G. Scarpetta for the fruitful discussions,
careful reading of the manuscript, 
and many useful suggestions.

\vspace{10. mm}

\begin{centerline}
{\bf REFERENCES}
\end{centerline}
\begin{enumerate}
\item\label{cobe}
R. Scaramella, N. Vittorio, \apj {\bf 353}, 372 (1990).\\
J.R. Gott, C. Park, R. Juszkiewicz, W.E. Bies, D.P. Bennet, F.R.
Bouchet, A. Stebbins, \apj {\bf 352}, 1 (1990).\\
{\it The Cosmic Microwave Background: 25 Years Later},
ed. N. Mandolesi, N. Vittorio, Kluwer, Dordrecht, (1990).\\
R. Scaramella, N. Vittorio, \apj {\bf 375}, 439 (1991).
\item\label{peebles}
P.J.E. Peebles, {\it Principles of Physical Cosmology},
Princeton Univ. Press, Princeton (1993).
\item\label{shane}
C.D. Shane, C.A. Wirtanen {\it Publ. Lick Obs.}, {\bf XXII}, Part I, 1  (1967).
\item\label{rc3}
G. de Vaucouleurs {\it et al.}, {\it RC3 Catalougue},
Springer--Verlag, Berlin (1991).
\item\label{groth}
E.J.Groth,  P.J.E. Peebles, \apj {\bf 217}, 385 (1977).
\item\label{loveday}
J. Loveday, G. Efstathiou, B.A. Peterson, S.J. Maddox, \apj {\bf 400}, L43
(1992).
\item\label{kirshner}
R.P. Kirshner, A. Oemler Jr., P. Schechter 
and S.A. Shectman, \apj {\bf 248}, L57 (1981).
\item\label{kennefick}
J.D. Kennefick, S.G. Djorgovki, and R.R. de Carvalho, {\it Astron. J.}
{\bf 110}, 2553 (1995).
\item\label{beks}
T.J. Broadhurst, R.S. Ellis, D.C. Koo, A.S. Szalay, {\it Nature}, {\bf 343},
726 (1990).
\item\label{bartelmann}
M. Bartelmann and P. Schneider, \aa {\bf 284}, 1 (1994).
\item\label{sakharov}
A. Sakharov, {\it Zh. Eksp. Teor. Fiz.}, {\bf 49}, 245 (1965).
\item\label{kolb}
Ya.B. Zel'dovich, I.D. Novikov, {\it Struttura ed evoluzione dell'Universo},
Ed. Riuniti, Roma (1985).\\
E.W. Kolb, M.S. Turner, {\it The Early Universe},
Ed. Addison--Wesley, New York (1990).
\item\label{guth}
A.A. Starobinsky, \pl {\bf B 91}, 99 (1980).\\
A.H. Guth, \pr {\bf D 23}, 347 (1981).\\
A. Albrecht, P.J. Steinhardt, \prl {\bf 48}, 1220 (1982).\\
A.D. Linde, \pl {\bf B 108}, 389 (1982).\\
A. D. Linde,  \pl {\bf B 114}, 431 (1982).\\
A.D. Linde,  \pl {\bf B 129}, 177 (1983).\\
D.S. Goldwirth and T. Piran {\it Phys Rep.}, {\bf 214}, 223 (1992).\\
F. Lucchin and S. Matarrese, \pr {\bf D 32}, 1316 (1985).\\
R.H. Brandenberger, \rmp {\bf 57}, 1  (1985).
\item\label{caianiello}
E.R. Caianiello, M. Gasperini, and G. Scarpetta, \cqg {\bf 8}, 659 (1991).
\item\label{la}
D. La, P.J. Steinhardt, \prl {\bf 62}, 376 (1989).\\
P.J. Steinhardt, F.S. Accetta, \prl {\bf 64}, 2740 (1990).
\item\label{birrell}
N.D. Birrell, P.C.W. Davies,  {\it Quantum Fields in Curved Space}
Cambridge Univ. Press, Cambridge (1982).
\item \label{adler}
D.W. Sciama, {\it Mon. Not. R. Astrom. Soc.} {\bf 113}, 34 (1953). \\
C. Brans  and R.H. Dicke,  {\it Phys. Rev} {\bf 124}, 925 (1961). \\
A. Zee, {\it Phys. Rev. Lett.} {\bf 42}, 417 (1979). \\
L. Smolin,  {\it Nucl. Phys.}  {\bf B 160}, 253 (1979).  \\
S. Adler, {\it Phys. Rev. Lett.} {\bf 44}, 1567 (1980).
\item\label{scherk}
J. Scherk and H.J. Schwarz, \np {\bf B 81}, 118 (1974).
\item \label{veneziano}
G. Veneziano \pl {\bf B 265},  287 (1991).\\
M. Gasperini, J. Maharana and G. Veneziano \pl {\bf B 272},  277 (1991).\\
K.A. Meissner and G. Veneziano \pl {\bf B 267}, 33 (1991).\\
M. Gasperini and G. Veneziano, \modpl {\bf 8 A}, 3701 (1993).\\
M. Gasperini and G. Veneziano, {\it Astropart. Phys.} {\bf 1}, 317 (1993).\\
M. Gasperini and G. Veneziano, \pl {\bf 329 B}, 429 (1994).
\item\label{dealwis},
K. Kikkawa, M. Yamsaki, \pl {\bf B 149}, 357 (1984).\\
S. P. de Alwis, \pl {\bf B 168}, 59 (1986).
\item\label{green}
M. Green, J. Schwarz, E. Witten, {\it Superstring Theory},
Cambridge Univ. Press, Cambridge (1987).
\item \label{tseytlin}
A.A. Tseytlin, \ijmp {\bf A 4}, 1257 (1989).\\
A.A. Tseytlin and C. Vafa \np {\bf B 372}, 443 (1992).
\item\label{pls} 
S. Capozziello, R. de Ritis, C. Rubano,
\pl {\bf 177 A}, 8 (1993).
\item\label{wdwstringhe} 
S. Capozziello, R. de Ritis,
 \ijmp {\bf 2 D}, 373 (1993).
\item\label{ndim} 
S. Capozziello, R. de Ritis, P. Scudellaro,
\ijmp {\bf D 2}, 463 (1993).
\item\label{cimento}
S. Capozziello, R. de Ritis, C. Rubano, and P. Scudellaro,
{\it La Rivista del Nuovo Cimento} {\bf 4}, 1 (1996). 
\item\label{dual}  
S. Capozziello, R. de Ritis,
\ijmp {\bf 2 D}, 367 (1993).
\item\label{copeland}
E.J. Copeland, A. Lahiri, and D. Wands, \pr {\bf D 50}, 4868 (1994).
\item\label{weinberg}
S. Weinberg, \rmp {\bf 61}, 1 (1989).
\item\label{qua} 
S. Capozziello, R. de Ritis, C. Rubano, and P. Scudellaro
 \ijmp {\bf 4D}, 767 (1995). 
\item\label{bra} 
S. Capozziello, R. de Ritis, C. Rubano, and P. Scudellaro
 \ijmp {\bf 5D}, 85 (1996). 
\item\label{any} 
S. Capozziello and R. de Ritis, \ijmp {\bf 5D}, 209 (1996). 
\item\label{lambdat}
S. Capozziello and R. de Ritis,  \grg {\bf 29}, 1425  (1997).
\item\label{damour}
T. Damour and A.M. Polyakov, \np {\bf 423 B}, 532 (1994).
\item\label{easther}
R. Easther and K. Maeda, {\it hep-th}/9509074 (1995).
\item\label{matter}
S. Capozziello and R. de Ritis, \pl {\bf 195 A}, 48 (1995).
\item\label{asy}
S. Capozziello and R. de Ritis, \pl {\bf  208 A}, 181 (1995).
\item\label{dickey}
T. Damour, G.W. Gibbons, and J.H. Taylor, \prl {\bf 61}, 1151 (1988).\\
C.M. Will, {\it Theory and Experiments in Gravitational Physics},
Cambridge Univ. Press, Cambridge (1993).\\
J.O. Dickey {\it et al.} {\it Science} {\bf 265}, 482 (1994).

\end{enumerate}

\newpage

\vspace{5. mm}

\begin{center}
\begin{tabular}{|l|l|l|r|}\hline\hline
      $\Lambda\neq 0$ & $k\neq 0$ & $D\neq 0$ \\ \cline{1-3}
      $\Lambda\neq 0$ & $k\neq 0$ & $D=0 $ \\ \cline{1-3}
      $\Lambda\neq 0$ & $k=0 $    & $D=0$ \\ \cline{1-3}
      $\Lambda=0$     & $k\neq 0$ & $D\neq 0$\\ \cline{1-3}
      $\Lambda=0$     & $k=0$ & $D\neq 0$\\ \cline{1-3}
      $\Lambda=0$     & $k=0$     & $D=0$\\ \cline{1-3}
      $\Lambda=0$     & $k\neq 0$ & $D=0$\\ \cline{1-3}
      $\Lambda\neq 0$ & $k=0$ & $D\neq 0$\\ \hline\hline
\end{tabular}
\end{center}  

 \vspace{2. mm}

{\bf Tab.1}: {\it The possible combinations of $\Lambda$,  $k$ and $D$.
We are considering any $\gam$ and $n$. In a lot of cases these two
parameters are related. For $\Lambda\neq 0$, we mean positive
and negative values; for $k\neq 0$, we mean $k=\pm 1$.}

\vspace{10. mm}

\begin{center}
\begin{tabular}{|l|l|l|l|r|}\hline\hline
  $\gam=1$   & $\Lambda\neq 0$ & $k=\pm 1,0$ & $n=1$ \\ \cline{1-4}
  $\gam=1$   & $\Lambda= 0$ & $k=\pm 1,0$ & $n=1$ \\ \cline{1-4}
  $\gam=1$   & $\Lambda\neq 0$ & $k=0 $ & $n\neq 1$ \\ \cline{1-4}
  $\gam=1$   & $\Lambda\neq 0$ & $k\neq 0$ & $n\neq 1$\\ \cline{1-4}
  $\gam=1$   & $\Lambda=0$     & $k=0$ & $n\neq 1$\\ \cline{1-4}
  $\gam=1$   & $\Lambda=0$     & $k\neq 0$ & $n\neq 1$\\ \hline\hline
\end{tabular}
\end{center}  

\vspace{2. mm}

{\bf Tab.2}: {\it For $\gam=1$,  we  have particularly
interesting cases which we summarize here.  $D$ is different 
from zero and positive.}

\vfill

\end{document}